# The physics of the longitudinal light clock

**Giovanni Zanella**

*Studioso Senior dello Studium Patavinum*
*Università di Padova, Italy*
*giovanni.zanella@unipd.it*

___________________________________________________________

**Abstract**

*The standard analysis of the behavior of the longitudinal light clock, in inertial motion, is not convincing in order to respect the postulates of the Special Relativity. Indeed, the time necessary to the light to travel from a mirror to the other in the moving clock is the same for a comoving observer, while it appears asymmetrical to a stationary observer. On the contrary, such asymmetry can be restored if the Doppler effect of the light emitted from a moving source is extended also to space and time.*
*As a consequence, the known Lorentz's contraction of the bodies in inertial motion is put in discussion.*

___________________________________________________________

**1. Introduction**

The standard analysis of the behavior of a longitudinal light clock (LLC) in uniform translatory motion (Fig.1), using the methods of the theory of the Special Relativity (SR), shows an asymmetry which is not in accord with the two postulatees of SR. We recall here these postulates as expressed by A. Einstein [1]:

1. *The laws by which the states of physical systems undergo change are not affected, whether these changes of state be referred to the one or the other of two systems of co-ordinates in uniform translatory motion.*
2. *Any ray of light moves in the "stationary" system of co-ordinates with the determined velocity c, whether the ray be emitted by a stationary or by a moving body.*

According with the *principle of relativity*, the *first postulate* tells us that an observer within a Lab in uniform motion of translation with respect to a stationary reference, cannot with experiments internal to the Lab to discover if the Lab is moving, or not. On the other hand, an observer at rest in the stationary reference cannot discern the motion of the Lab only looking to the physical phenomena which appear inside the Lab itself.

The *second principle* tells us, differently from the Galileo's view, that the motion of a source of light cannot be imparted to the emitted light from the source itself.

As concerns the time, it can be intended as the path of a ray of light along a graduated rigid axis. Indeed, being A and B two points of the space where are placed two clocks, Einstein wrote [1]:

*… we establish by definition that the time required by light to travel from A to B equals the time it requires to travel from B to A. Let a ray of light start at the "A time" $t_A$ from A towards B, let at the "B time" $t_B$ be reflected at B in the direction of A, and arrive again at A at time "A time" $t'_A$. … In agreement with experience we further assume the quantity*

$$\frac{2AB}{t'_A - t_A} = c \qquad (1)$$

*to be a universal constant: the velocity of light in empty space."*
In accordance with definition the two clocks synchronize if

$$t_B - t_A = t_{A'} - t_B \qquad . \qquad (2)$$

Obviously, with this procedure it is the rate of the two clocks that is tuned and not the start point of the time marked by them.

Therefore, $t'_A - t_A$ of Eq.(1) expresses the time spent by the light ray to travel over the path A-B-A, so $(t'_A - t_A)/2$ expresses indifferently the time spent by the light ray over the path AB or BA. Thus, if conventionally we put $c=1$, the path of a ray light represents too the time necessary to the light to cross the path itself, in any inertial reference.

Starting from these premises, we analyze the behavior of the moving LLC showing the necessity of an extension of the Doppler effect also to space and time involved by the moving light source.

In the following analysis the observer is supposed ubiquitous in the space ascribed to his reference system, as also Einstein intended.

Besides, we intend the LLC to have a mass so high to neglect recoil effects due to the reflections of the ray of light on the two mirrors.

## 2. Galilean view

Fig.2 shows the schema of the LLC in uniform translatory motion (or inertial motion) with velocity *u* with respect to a stationary reference, in a Galilean scenario.

The Galilean scenario intends time and space to be absolute and besides it considers the composition of the velocity of the light with that of the LLC.

Therefore, the time $T_{AB'}$ necessary to the ray of light to travel from end A to end B, as viewed from the stationary reference, will be

$$T_{AB'} = \frac{L}{c+u} + \frac{uT_{AB'}}{c+u} \quad , \quad \text{that is}$$

$$T_{AB'} = \frac{L}{c} \qquad (3)$$

where *L* is the proper distance of the mirrors, or the proper length of the rod of support.

At the same manner, the time $T_{B'A''}$ necessary to the ray of light to travel from B to A, will be

$$T_{B'A''} = \frac{L}{c-u} - \frac{uT_{B'A''}}{c-u} \quad , \quad \text{that is}$$

$$T_{B'A''} = \frac{L}{c} \qquad (4)$$

Therefore, the *first principle* of SR is respected in the Galilean scenario of LLC, because looking to the internal behavior of LLC from the stationary it is not possible to discern the motion of the clock itself.

Obviously, the same result appears when the ray of light is replaced by a small sphere of negligible mass which is reflected elastically between the mirrors.

## 3. The standard analysis of LLC

Introducing the *second principle* of SR, that is the invariance of the velocity *c* of the light in the empty space, the time $T_{AB'}$ required to a light ray to cross the path AB' (Fig.3), supposing $u << c$ (*Lorentz's factor* $\gamma \cong 1$), would be

$$T_{AB'} = \frac{L}{c} + \frac{uT_{AB'}}{c}, \quad \text{that is}$$

$$T_{AB'} = \frac{L}{c-u}, \tag{5}$$

while for the path B'A''

$$T_{B'A''} = \frac{L}{c} - \frac{uT_{B'A''}}{c}, \quad \text{that is}$$

$$T_{B'A''} = \frac{L}{c+u}. \tag{6}$$

We can note that also Einstein arrived to the same result of Eq.(5) and Eq.(6) by a thought experiment [1]. He, used rays of light and placed two synchronized clocks (with the same start of the time) at the ends A and B of a rigid rod. So, imparted to the rod a uniform motion of parallel translation with velocity *u*, along the direction of its axis, he wrote [1]:

*Let a ray of light depart from A at the time $t_A$, let it be reflected at B at the time $t_B$, and reach A again at the time $t_{A'}$. Taking in consideration the principle of the constancy of the velocity of light we find*

$$t_B - t_A = \frac{L}{c-u} \quad \text{and} \quad t_{A'} - t_A = \frac{L}{c+u} \tag{7}$$

*where L denotes the length of the moving rod, measured in the stationary system.*
*Observers moving with the moving rod would thus find that the two clocks were not synchronous, while observers in the stationary system would declare the clocks to be synchronous.*

Here the mistake of Einstein appears evident, because a comoving observer with the rod cannot have means to detect its inertial motion. Subsequently he repaired the mistake pointing out that the asymmetry was detected only by the stationary observer looking to the moving rod [2].
We known, instead, that this asymmetry is still against the *first postulate* of SR, as well as the composition of the velocity of the light with the velocity *u* is against the *second postulate* of SR.

Therefore, if the standard analysis of LLC doesn't conduct to a result physically acceptable it appears evident that further underlying physics must be explored.

As we will see, a further crack of the Einstein's view will appear when the known *Lorentz's contraction* of the bodies in inertial motion goes in contrast with the dilation of the time [3].

## 4. Physics of LLC

Suppose that S and S' are two Cartesian reference systems having parallel axes, where the *x'*-axis is in uniform translation with velocity $u << c$ ($\gamma \cong 1$) along the positive *x*-axis (Fig.4). The time started when the origin O of S and the origin O' of S' were in coincidence ($t = 0$).

Fig.4 shows the snapshot, at the time *t*, of the two–dimension wave-front of light generated in the time *t=0*, and in the empty space, from a isotropic source located in the coincidence point of the origins of S' and S.

Now, the wave-front of Fig.4 is necessarily a concentric sphere with O', if viewed from a comoving observer with S'. This eventuality is possible only if the *Doppler effect* involved by the motion of O' is extended also to the space and consequently to the time. In practice, the space becomes compressed in front to O', in the direction of *u*, and dilated in the opposite direction, as viewed from the stationary reference [3].

Therefore, with respect to O', as viewed from S, two paths are associated to the wave-front which propagates along the *x*-axis (solid arrows of Fig.4): the path $X_f$ corresponding to the forward path along the direction of the motion of S' and the path $X_b$ ascribed to the opposite direction.

So, the path $X_f$ results a contraction of $X=ct$, that is (Fig.4)

$$X_f = X - ut = X\left(1 - \frac{u}{c}\right) = X\left(\frac{c-u}{c}\right), \qquad (8)$$

and the path $X_b$ results a dilation of $X=ct$, that is (Fig.4)

$$X_b = X + ut = X\left(1 + \frac{u}{c}\right) = X\left(\frac{c+u}{c}\right), \qquad (9)$$

Hence, looking to Eq,(5) and Eq.(6), we must contract *L* with the factor $(c-u)/c$, when the light ray travels forward from A to B' and dilate the space with the factor $(c+u)/c$, when the light travels back from B' to A''. So, the times $T_{AB'}$ and $T_{B'A''}$ become

$$T'_{AB'} = \frac{L}{c-u}\left(\frac{c-u}{c}\right) = \frac{L}{c} \quad \text{and} \quad T'_{B'A''} = \frac{L}{c+u}\left(\frac{c+u}{c}\right) = \frac{L}{c} \quad . \tag{10}$$

In practice, we have obtained $T'_{AB'} = T'_{B'A''}$, independently on the velocity $u$, and also the constancy of the velocity $c$ of the light in either directions of motion (Fig.5).
On the other hand, introducing the known *Lorentz's dilation* on $T'_{AB'}$ and $T'_{B'A''}$ [1], we have

$$T_{AB'}/\sqrt{1-u^2/c^2} = T_{B'A''}/\sqrt{1-u^2/c^2} = \frac{L}{c}/\sqrt{1-u^2/c^2} \quad . \tag{11}$$

Eq.(11) compels us to dilate $L$ in $L' = L/\sqrt{1-u^2/c^2}$, in accord with the notion of the time intended as the path of a light ray (Fig.5).
It is remarkable that in this relativistic view a same scale factor (the *Lorentz's factor* $1/\sqrt{1-u^2/c^2}$) pertains the whole moving system so that the observer, comoving with the clock, has not means to discern its inertial motion.
In lack of the *Doppler effect* on space and time, we would find as overall time for the light ray, to cover a forward and back path, the time

$$T' = \frac{L}{c-u} + \frac{L}{c+u} = \frac{2cL}{c^2-u^2} \quad , \tag{12}$$

so $L = \frac{cT'}{2}\left(1-\frac{u^2}{c^2}\right)$, but for SR $T' = \frac{T}{\sqrt{1-\frac{u^2}{c^2}}}$, and then we would have a

shortened length of LLC, that is

$$L\sqrt{1-\frac{u^2}{c^2}} \quad . \tag{13}$$

7. **Conclusions**

The standard analysis of the behavior of the longitudinal light clock, in uniform translatory motion with respect to a stationary reference, reveals an asymmetry of the time against the *first principle* of Special Relativity [1]. Indeed, the time necessary to a light ray to travel from a mirror to the other would result dilated in the forward path, along the direction of the motion of the clock, and shortened in the back path. As a consequence, it would be possible to discern the inertial motion of the clock from a stationary reference, looking to the internal behavior of the clock itself.

The paper shows instead that such asymmetry does not exist, because it can be restored if the *Doppler effect* is extended also to the space and to the time involved by the moving light source [3].

At least, introducing the known relativistic time dilation, by the *Lorentz's factor*, also the length of the longitudinal light clock must dilate at the same manner, against the known *Lorentz's contraction*.

**Figure captions**

Fig.1. Schematic representation of a *longitudinal light clock* (LLC).
Fig.2. Galilean view of the behavior of LLC in uniform translatory motion with velocity $u$ with respect to a stationary reference (see text).
Fig.3. Schematic representation of LLC of Fig.1 in the standard scenario with $u \ll c$ (see text).
Fig.4. Two-dimensional representation of the wave-front of light emitted isotropically from a source in uniform translatory motion the empty space as viewed from a stationary reference S. The source, located in the origin of the system S', moves with velocity $u \cong 0.25\ c$ ( *Lorentz's factor* $\gamma \cong 1$) with respect to system S (see text).
Fig.5. Schematic representation of LLC of Fig.1 in a relativistic scenario, where $\gamma$ represents the *Lorentz's factor* (see text).

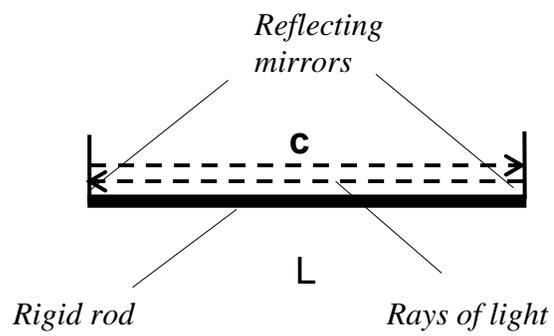

**Fig. 1**

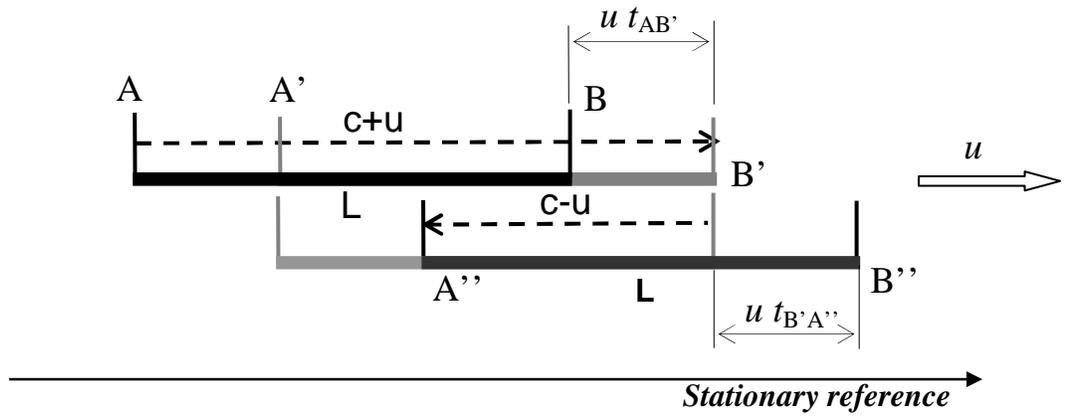

**Fig. 2**

**Fig. 3**

**Fig. 4**

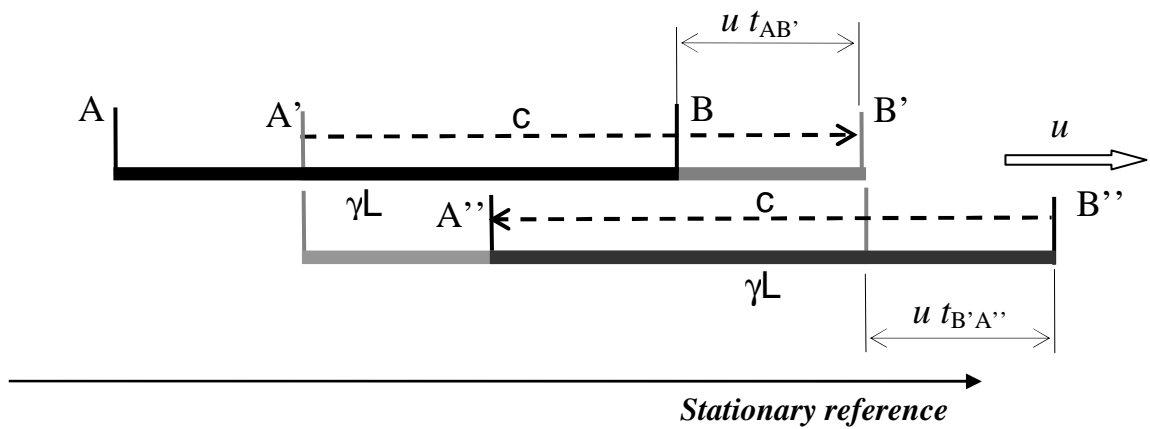

**Fig. 5**